\documentclass[conference]{IEEEtran}
\usepackage{cite}
\usepackage{amsmath,amssymb,amsfonts}
\usepackage{algorithmic}
\usepackage{graphicx}
\usepackage{textcomp}
\usepackage[table, svgnames, dvipsnames]{xcolor}
\usepackage{hyperref}
\usepackage[frozencache=true,cachedir=minted]{minted}

\makeatletter
\let\@float@c@listing\@caption
\makeatother

\begin{document}

\title{Enhancing Quantum Software Development Process with Experiment Tracking}

\author{
\IEEEauthorblockN{Mahee Gamage}
\IEEEauthorblockA{University of Jyväskylä\\
Jyväskylä, Finland\\
mahee.s.hewagamage@jyu.fi
}
\and
\IEEEauthorblockN{Otso Kinanen}
\IEEEauthorblockA{University of Jyväskylä\\
Jyväskylä, Finland\\
otso.j.r.kinanen@jyu.fi
}
\and
\IEEEauthorblockN{Jake Muff}
\IEEEauthorblockA{Quantum Algorithms and Software\\
VTT Technical Research Centre of Finland\\
Espoo, Finland\\
jake.muff@vtt.fi}
\and
\IEEEauthorblockN{Vlad Stirbu}
\IEEEauthorblockA{University of Jyväskylä\\
Jyväskylä, Finland\\
vlad.a.stirbu@jyu.fi}
}

\maketitle

\begin{abstract}
As quantum computing advances from theoretical promise to experimental reality, the need for rigorous experiment tracking becomes critical. Drawing inspiration from best practices in machine learning (ML) and artificial intelligence (AI), we argue that reproducibility, scalability, and collaboration in quantum research can benefit significantly from structured tracking workflows. This paper explores the application of MLflow in quantum research, illustrating how it enables better development practices, experiment reproducibility, decision making, and cross-domain integration in an increasingly hybrid classical-quantum landscape.
\end{abstract}

\IEEEpeerreviewmaketitle

\section{Introduction}

Quantum software engineering is an emerging discipline fueled by recent advances in quantum hardware. Despite being part of the broader field of software engineering, quantum software engineering has to take into account the current and near-term technical limitations of the Noisy intermediate-scale quantum (NISQ) \cite{Preskill2018quantumcomputingin} era hardware. The limited number of qubits available in quantum processing units (QPUs) and their quality have a direct impact on the complexity of the programs that can be executed, highlighting important challenges \cite{murillo2025challenges}, which have to be addressed to effectively leverage the promises of quantum computing.

A quantum software or algorithm developer must navigate this complicated environment in a structured way following a well-defined software development lifecycle \cite{qsdlc}, supported by specialized tools \cite{Kinanen2025}. Given the limited availability of quantum hardware, the development process begins on quantum simulators. Subsequently, as the program or algorithm matures, its execution is performed on the actual hardware. Due to the unique nature of quantum environments, developers must collect relevant data to guide each development stage. Early on, the focus is on the program or algorithm itself, but once executed on a QPU, data collection expands to include hardware-specific information such as calibration and qubit quality.

In this paper, we propose a quantum experiment tracking system built on MLflow, an open-source platform widely adopted in ML/AI development. By leveraging its’s existing capabilities, we focus on developing quantum-specific extensions while benefiting from its mature ecosystem and the operational expertise of an already skilled workforce.

\section{Background and Motivation}

Experiment tracking \cite{zaharia2018accelerating} was introduced as a tool that addresses the challenges faced by machine learning (ML) developers in the following four areas: multitude of tools, experiment tracking, reproducibility, and production deployment. We can observe that quantum software development has some similarities with ML/AI development. For example, developers have the option to use several quantum software development toolkits (e.g. Qiskit\footnote{\href{https://www.ibm.com/quantum/qiskit}{https://www.ibm.com/quantum/qiskit}}, PennyLane\footnote{\href{https://pennylane.ai}{https://pennylane.ai}}, Qrisp\footnote{\href{https://www.qrisp.eu}{https://www.qrisp.eu}}, etc.) that enables them to interact with a multitude of hardware vendors in order to execute their quantum routines. During this process, developers have to collect a plethora of software and hardware related data to steer the software development activities in the right direction.

\begin{figure}
    \centering
    \includegraphics[width=\linewidth]{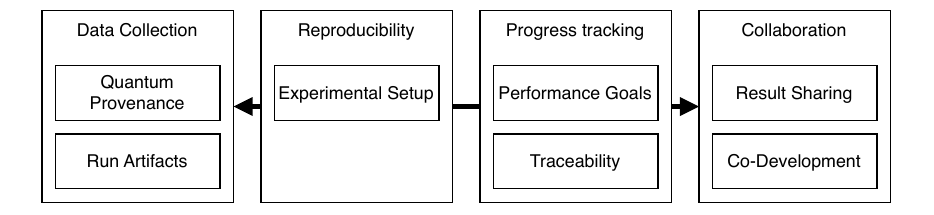}
    \caption{Quantum software development activities supported by experiment tracking}
    \label{fig:activities}
\end{figure}

Experiment tracking fundamentally involves collecting and logging data about the software and hardware used. QProv \cite{qprov} introduces a quantum provenance model with a data schema that captures four key categories: quantum circuit, quantum computer, compilation, and execution. These form the core of experiment tracking and can be extended with application-specific artifacts. Given the experimental nature of current quantum hardware, run data is essential not only for evaluating performance but also for ensuring reproducibility, since hardware reliability can vary between runs, and simulators cannot fully replace real hardware experimentation \cite{senapati2023reproducibility}. Additionally, the collected data supports progress tracking by providing a structured view of how performance goals are met, helping guide development to subsequent stages \cite{Kinanen2025}. These artifacts also offer traceability and support decision-making. Finally, since quantum software development is inherently collaborative, experiment tracking enhances result sharing and supports joint development between hardware and software teams, as depicted in Fig. \ref{fig:activities}.

\section{Quantum Experiment Tracking}

\begin{figure}
    \centering
    \includegraphics[width=\linewidth]{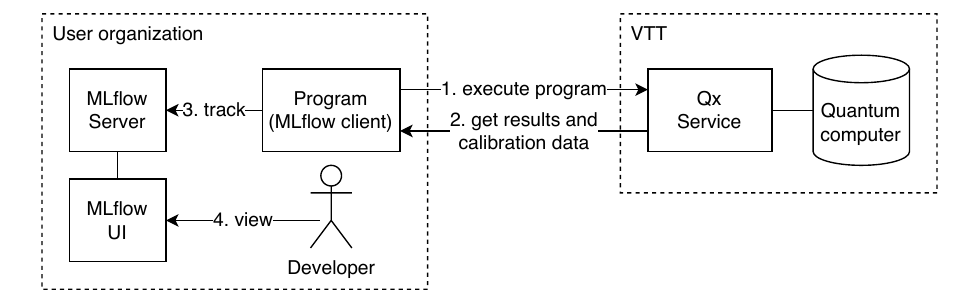}
    \caption{Experiment setup: the user executes a program on the on a quantum computer exposed via the VTT QX service and tracks the experiment results to the MLflow tracking server operated by its own organization}
    \label{fig:experiment}
\end{figure}

Since its introduction, MLflow\footnote{\href{https://mlflow.org}{https://mlflow.org}} has become a standard tool for managing the development lifecycle of ML projects, including more complex domains like Generative AI and Large Language Models. In our study, we used MLflow to track quantum experiment data from programs executed on an IQM 50-qubit quantum computer, operated by VTT via the QX service. The user's organization hosts the MLflow tracking server and UI, while data is logged using the MLflow client integrated into the quantum program. The experiment setup is illustrated in Fig.~\ref{fig:experiment}.

\begin{listing}
    \fontsize{6.4}{8}
    \begin{minted}[linenos=true,highlightlines={2,4,9,10,13,16}]{python}
import mlflow
mlflow.set_experiment("Qx VTT Demo for QCE")
with mlflow.start_run():
    mlflow.set_tag("Training info", "Qiskit on Qx")
    provider = IQMProvider("https://qx.vtt.fi/api/devices/q50")
    backend = provider.get_backend()
    shots = 500
    result = demo_function(backend, shots)
    mlflow.log_param("shots", shots)
    mlflow.log_figure(plot_histogram(result.get_counts()),
                                     "results.png")
    calibration_set_id = str(result.results[0].calibration_set_id)
    mlflow.log_text(calibration_set_id, "calibration_set_id.txt")
    calibration_data = get_calibration_data(backend.client,
                                            calibration_set_id)
    mlflow.log_dict(calibration_data, "calibration_data.json")
    \end{minted}
    \caption{Tracking experiment data with MLflow. The implementation of \texttt{demo\_function}, \texttt{plot\_histogram} and \texttt{get\_calibration\_data} functions has been omitted for brevity.}
    \label{lst:logging}
\end{listing}

\textbf{Experiment tracking} data collection is performed using the MLflow client library that is included in the quantum program. MLflow client can name experiments, set tags on individual runs, and log parameters, metrics, or free-form artifacts (e.g. figures, datasets, etc), capabilities that can capture the QProv attributes. A simplified\footnote{Full program available at \href{https://github.com/qubernetes-dev/q8s-examples}{https://github.com/qubernetes-dev/q8s-examples}} program is depicted in the Listing. \ref{lst:logging}.

MLflow client library can be used to process \textbf{data from multiple experiments} using the search functionality. The result of a query can be used with other popular libraries to create new artifacts that support decision making activities. A simplified program is depicted in the Listing. \ref{lst:retrieval}.

\begin{listing}
\fontsize{6.4}{8}
\begin{minted}[linenos=true]{python}
import mlflow
import pandas as pd
import matplotlib.pyplot as plt
experiment_name = "Qx VTT Demo for QCE"
exp = mlflow.get_experiment_by_name(experiment_name)
runs_df = mlflow.search_runs(experiment_ids=[exp.experiment_id])
...
# Data-frame can be further processed with pandas and matplotlib
\end{minted}
\caption{Processing experiments data with MLflow}
\label{lst:retrieval}
\end{listing}

\section{Discussion and Future Work}

In this paper, we demonstrated how MLflow can serve as a foundation for quantum experiment tracking. We showed that MLflow is well-suited to quantum research, supporting improved development practices, reproducibility, informed decision-making, and collaboration. Instead of building standalone solutions like the QProv provenance system \cite{qprov} or Aqueduct\footnote{\href{https://github.com/AqueductHub}{https://github.com/AqueductHub}}, we advocate for leveraging mature and widely adopted tools with strong community and ecosystem support. As next steps, we plan to automate the collection of data according to QProv format, with selected QPU providers, and integrate them into quantum software development workflows.

\section*{Acknowledgment}

This work has been supported by the Academy of Finland (project DEQSE 349945), Business Finland (EM4QS 155/31/2024), Finnish Ministry of Education and Culture through the Quantum Doctoral Education Pilot Program (QDOC VN /3137/2024-OKM-4) and the Research Council of Finland through Finnish Quantum Flagship project (359240, JYU).

\bibliographystyle{abbrv}

\end{document}